\documentclass[%
preprint,
 amsmath,amssymb,
 aps,
]{revtex4-1}

\usepackage{graphicx}
\usepackage{dcolumn}
\usepackage{bm}

\newcommand{\rmR}{{\rm R}}

\newcommand{\rmI}{{\rm I}}

\newcommand{\bra}{\langle}
\newcommand{\ket}{\rangle}

\newcommand{\bd}{\begin{displaymath}}
\newcommand{\ed}{\end{displaymath}}

\newcommand{\bea}{\begin{eqnarray}}
\newcommand{\eea}{\end{eqnarray}}
\newcommand{\beas}{\begin{eqnarray*}}
\newcommand{\eeas}{\end{eqnarray*}}
\newcommand{\be}{\begin{equation}}
\newcommand{\ee}{\end{equation}}

\def\rmi{{\rm i}}

\def\vecR{\mbox{\boldmath $R$}}
\def\vecX{\mbox{\boldmath $X$}}

\def\vec0{\mbox{\boldmath $0$}}

\def\veceta{\mbox{\boldmath $\eta$}}
\def\vecxi{\mbox{\boldmath $\xi$}}
\def\vecx{\mbox{\boldmath $x$}}

\begin{document}
\title{
Correspondence between  phase oscillator network and  
classical XY model with the same infinite-range interaction in statics}

\author{Tatsuya Uezu$^1$}\email{uezu@ki-rin.phys.nara-wu.ac.jp} 
\author{Tomoyuki Kimoto$^2$} 
\author{Shuji Kiyokawa$^3$} 
\author{Masato Okada$^4$}
 \affiliation{
$^1$Graduate School of Humanities  and Sciences,
Nara Women's University, Nara 630-8506, Japan
}
\affiliation{%
$^2$Oita National College of Technology, Oita 870-0152, Japan
}%
\affiliation{
$^3$Faculty of Science,Nara Women's University, Nara 630-8506, Japan
}%
\affiliation{
$^3$Graduate School of Frontier Sciences, The University of Tokyo, Kashiwa, Chiba 277-8561, Japan
}
\date{\today}

\begin{abstract}
We study the phase oscillator networks with distributed natural frequencies  and 
classical XY models both of which have a class of infinite-range interactions in common.
We find that the integral kernel of the self-consistent equations (SCEs) for  oscillator networks
correspond to  that of the saddle point equations (SPEs) 
 for   XY models, and that the quenched randomness (distributed natural frequencies)
corresponds to thermal noise. We  find a sufficient condition that
 the probability density of natural frequency distributions
 is one-humped in order that 
the kernel in the oscillator network
is strictly decreasing as that in the XY model. 
Furthermore, 
 taking the uniform and Mexican-hat type interactions, we prove  the  one to one
 correspondence between the solutions of the SCEs and SPEs.
 As an application of the correspondence, we study the associative memory type
 interaction. In the XY model  with this interaction,
 there exists a peculiar one-parameter family of solutions.  For the oscillator network,
 we find a non-trivial solution, i.e., a limit cycle oscillation.
\end{abstract}
\maketitle


\newfont{\msamfnt}{msam10}
\newcommand{\gsim}{\mbox{$\;$ \msamfnt\symbol{'046} $\;$}}
\newcommand{\llsim}{\mbox{$\;$ \msamfnt\symbol{'056} $\;$}}

\def\veco{\mbox{\boldmath $w^o$}}
\def\vecx{\mbox{\boldmath $x$}}
\def\vecw{\mbox{\boldmath $w$}}
\def\veceta{\mbox{\boldmath $\eta$}}
\def\vecR{\mbox{\boldmath $R$}}




Synchronization phenomena prevail in nature\cite{saunders,biological.clocks} and 
have attracted many researchers.  Among others, 
 Winfree studied biological rhythms and introduced
 a  phase description\cite{winfree}. Later, Kuramoto proposed a seminar model of 
 synchronization-desynchronization transitions, the so called Kuramoto
 model\cite{kuramoto-1}. 
  Since then, a lot of studies on the Kuramoto model and its extensions 
have been done\cite{rev.mod.phys,after2005}. 
%
On the other hand, the classical XY models have been studied
intensively and extensively mainly for short range interactions\cite{domb.green}.
If the interactions are the same, 
 complex order parameters are also the same in both models.
 
In the course of the study of a phase oscillator network with an infinite-range 
interaction\cite{uezu.osc}, 
we investigated the classical XY model with the same interaction 
and found complex order parameters 
obey similar equations in  both models\cite{kimoto.xy}.\par 
It is obvious that the phase oscillator network with
 the uniform natural frequency is equivalent to the classical XY model 
with temperature 0 if the interaction is common in two models.
In this paper,  we report 
that even for the phase oscillator network with distributed natural frequencies and the 
 XY model with non-zero temperature, 
there exists some correspondence between them.
We treat a class of interactions for which the Hamiltonian is expressed by
order parameters, and  derive  the correspondence of 
probability density functions for phases, and that between 
the self-consistent equations (SCEs) for the phase oscillator network
 and  the saddle point equations (SPEs) for the XY model.
We also find   a sufficient condition for
the probability density of natural frequency distributions in order
that the precise correspondence holds in both models, and 
that the quenched randomness corresponds to thermal noise.
 Furthermore, we study 
 the uniform interaction and the  Mexican-hat type interaction on a circle,
and prove the one to one correspondence of the solutions in both models.
Finally, as an application, we study  the associative memory type interaction.  
For the XY model with this interaction, there exists a peculiar solution, 
i.e., one-parameter family of solutions which we call the continuous solution\cite{kimoto.etal}.
By the correspondence, we immediately obtain the SCEs for the oscillator network.
We theoretically and numerically study both models and find
the continuous solution changes to a noisy limit cycle oscillation in the oscillator network.\\
\underline{Phase oscillator network} \\
Let us consider $N$ phase oscillators.
Let $\phi_j '$ be the phase of the  $j$th oscillator, and assume that 
it obeys the following differential equation:
\begin{eqnarray}
\frac{d}{dt} \phi_j ' &=& \omega _j + \sum_k J_{jk}\sin(\phi_k '- \phi_j ').
\label{eq:evolution}
\end{eqnarray}
Here, $\omega_j$ is natural frequency and it is drawn from the
probability density $g(\omega)$. We assume that  $J_{jk}=J_{kj}$,  the mean value of $\omega$
is $\omega_0$, and
 $g(\omega)$ is symmetric with respect to $\omega_0$,
\bea
g(\omega_0+x)&=&g(\omega_0-x).
\label{sym.g}
\eea
We put $\phi_j = \phi_j ' -\omega_0$ and define  $A_j$ and $\alpha_j$ by
\bea
A_j e^{i \alpha_j} &=& \sum_k J_{jk} e ^{i \phi_k}.
\label{alpha}
\eea
Since we have an interest in  stationary states, we assume $A_j$ and 
$\alpha_j$ do not depend on time.
By defining $\psi_j=\phi_j  - \alpha_j$, the evolution equation becomes
\bea
d \psi_j /dt &=& \omega _j-\omega_0 - A_j \sin \psi_j.
\eea 
Let $\hat{n}(\phi', t,j)$ be the probability density of $\phi'$ for
 the $j$th oscillator  at time $t$.
Assuming stationary rotation of the probability
 density and defining $\hat{n}( \phi', t, j)
=\bar{n}(\phi' -\omega_0t,j) \equiv n(\psi, j)$, 
the continuity equation becomes 
\bea
\frac{\partial}{\partial t} n(\psi, j)
&=& -\frac{\partial}{\partial \psi}
\biggl( \bigl( \omega_j -\omega_0 +A_j \sin \psi   \bigr) n(\psi, j)\biggr).
\eea
Its stationary solution is
\bea
\bigl( \omega_j -\omega_0 +A_j \sin \psi_j   \bigr) n(\psi, j) &=&C_j,\\
n(\psi, j) &=& n_s (\psi, j) +n_{ds }(\psi, j),
\eea
where $n_s$ and $ n_{ds}$ are densities for the synchronized and desynchronized
 oscillators, respectively.
For the stable synchronized oscillators, we get 
\bea
 n_s (\psi, j)&=& g(\omega_0 + A_j \sin \psi)A_j \cos \psi, \ |\psi|<\pi/2.
\label{osi.ns}
\eea
\underline{XY model} \\
The classical XY spins are denoted by
 $\vecX_j=(\cos \phi_j,  \sin \phi_j), j=1,2,\cdots, N$.
  The Hamiltonian is given by
\begin{eqnarray}
H &=& - \sum_{j<k} J_{jk}\cos(\phi_j - \phi_k)
= - \frac{1}{2}\sum_{j, k} J_{jk}\cos(\phi_j - \phi_k) +C,
\end{eqnarray}
where $C=\sum_j J_{jj}/2$.
The equilibrium state is described by the canonical distribution,
$P_{eq} = e^{- \beta H}/Z$ where $Z$ is the partition function,
 $\beta=1/T$ and $T$ is
 the temperature.  We put $k_B =1$.\\
\underline{Interaction and Order parameters} \\
We consider the following interaction:
\bea
J_{jk}&=&\frac{1}{N}\sum_{l=1}^L a_l q_{l,j} q_{l,k}, 
\label{interaction}
\eea
where $a_l>0$ and $q_{l,j}$ are real numbers.
We define the order parameters as
\bea
Q_{l}e^{i \Phi_{l}}&=& Q_{l, \rmR}+ i Q_{l, \rmI} =
\frac{1}{N} \sum_j q_{l,j}e^{i \phi_j}, \ l=1, \cdots, L.
\eea
Therefore, in the XY model,  the Hamiltonian is expressed as
\bea
H&=&- \frac{N}{2} \sum_{l=1}^L a_l (Q_{l, \rmR} ^2 + Q_{l, \rmI}^2) +C.
\label{beta}
\eea
By using the saddle point method, we obtain the partition function and
the probability density function $n(\phi,j)$ of $\phi_j$ for the $j$th spin as
\bea
Z &\propto & \exp[N(-\frac{\beta}{2} \sum_{l=1}^L a_l Q_l ^2 + \frac{1}{N} \sum_j \Omega_j)],\\
&& \exp[\Omega_j]= 
\int d \phi_j \exp[\beta \sum_l  a_l    q_{l,j}(Q_{l, \rmR} \cos \phi_j +Q_{l, \rmI} \sin \phi_j)]
\nonumber \\
&&= 2 \pi I_0 (\beta \Xi_j) \\
&& \Xi_j =   \sqrt{ (\sum_l  a_l q_{l,j}Q_{l, \rmR})^2+ (\sum_l a_l q_{l,j} Q_{l, \rmI})^2},
\label{def.xi}\\
&& \Xi_j \cos \phi_j ^0 = \sum_l  a_l    q_{l,j} Q_{l, \rmR}, 
\ \Xi_j \sin \phi_j ^0 =  \sum_l a_l    q_{l,j} Q_{l, \rmI}.
\\
n(\phi,j)&=&  \frac{\exp[\beta \Xi_j \cos (\phi-\phi_j ^0)]} {2 \pi I_0 (\beta \Xi_j)}.
\label{xy.n}
\eea
Here, $I_n(x)$ is the $n$th order modified Bessel function,
\bea
I_n(x)&=& \frac{1}{\pi} \int_0 ^{\pi} d \theta \exp[x \cos \theta] \cos(n\theta).
\eea
$n(\phi,j)$  is the so called von Mises distribution. This function corresponds to  (\ref{osi.ns}).
The SPEs are
\bea 
Q_l e^{i \Phi_l} &=& \frac{1}{N}\sum_j  
\int d \phi_j 
\exp[-\Omega_j+\beta \sum_{l'} q_{l',j} 
 (Q_{l', \rmR} \cos \phi_j + Q_{l', \rmI} \sin \phi_j)] q_{l,j}e^{i \phi_j} \nonumber\\
&=&  \frac{1}{N}\sum_j \frac{I_1 (\beta \Xi_j)} {I_0 ( \beta \Xi _j)} q_{l,j}e^{i  \phi_j ^0}. 
\label{spe}
\eea
Furthermore, we obtain the following relation from Eqs. (\ref{alpha}) and (\ref{interaction}):
\bea
A_j e^{i \alpha_j} &=&\Xi_j e^{i \phi_j ^0}.
\label{AZ}
\eea
In the oscillator network, the SCEs are
\bea
Q_l e^{i \Phi_l} 
&=& \frac{1}{N}\sum_j  2 \int_0 ^{\pi/2} d \psi 
g(\omega_0+A_j \sin \psi )A_j \cos^2 \psi \  q_{l,j}  e^{i \alpha_j}.
\label{sce}
\eea
Due to the assumption (\ref{sym.g}),  the desynchronized solutions
do not contribute to the order parameters.\\
\underline{Correspondence of integration kernels and that of randomness} \\
Let us define the following functions and coefficients:
\bea
u(x)& \equiv &I_1(x)/(x I_0(x)),
\label{def.u}\\
\bar{g}_{\omega_0,\sigma}(x)& \equiv &
2  \int_0 ^{\pi/2} d \psi g(\omega_0+ x \sin \psi)\cos^2 \psi,
\label{def.gbar}\\
\bar{u}_\beta(x) & \equiv & \beta u(\beta x).
\label{def.ubar}
\eea
Using these functions and Eq. (\ref{AZ}), SPEs  (\ref{spe})  and SCEs (\ref{sce}) are
rewritten as
\bea
Q_l e^{ i \Phi_l} &=& \frac{1}{N} \sum_j A_j \bar{g}_{\omega_0,\sigma}(A_j) q_{l,j} e^{\alpha_j},
\label{sce.barg}\\
Q_l e^{ i \Phi_l} &=& \frac{1}{N} \sum_j A_j \bar{u}_\beta (A_j) q_{l,j} e^{\alpha_j}.
\label{spe.baru}
\eea
From these equations, we find $ \bar{g}_{\omega_0,\sigma}(x)$ 
and $ \bar{u}_\beta (x)$ correspond. 
If we derive the concrete equations for order parameters in one model, 
we immediately obtain them in the other model. 
 We call these functions  the integration kernels 
because  these equations become integration equations in some cases 
as seen later.
Furthermore, from the value of the kernels at $x=0$,
 we have the following correspondence:
\bea
&& T \Longleftrightarrow  1/(\pi g(\omega_0))(=  \sqrt{2/\pi}\sigma),
\label{tvssigma} 
\eea
where the expression in the parentheses is for the Gaussian distribution, 
and $\sigma$ is the standard deviation of the natural frequency $\omega$.
The correspondence (\ref{tvssigma}) is also derived by comparing
 the phase transition points in both models.
 Equation (\ref{tvssigma}) 
 implies that the temperature corresponds to the width of distribution of
the natural frequency around the center $\omega_0$,
 that is,  thermal noise corresponds to the quenched randomness.\\
{\bf Sufficient condition that both kernels have the same property}\\
$ \bar{u}_{\beta}(x) $ and $\bar{g}_{\omega_0,\sigma}(x)$
 take a finite value  at $x=0$, and tend to 0 as $x$ tends to $\infty$.
In addition to these properties, $  \bar{u}_{\beta}(x) $ has the following property:
\bea
&& \frac{d \bar{u}_\beta}{d x}(x) <0, \mbox{ for $x>0$.} 
\label{u3}
\eea
$g'(\omega)<0$  for $\omega >\omega_0$ is a sufficient condition 
 for the property (\ref{u3}),
 that is 
$g(\omega)$ has a single maximum at $\omega_0$
 and is strictly decreasing for $\omega >\omega_0$.
Hereafter, we assume this property for $g(\omega)$. 
 By using these properties, we prove  the correspondence of solutions 
in the followings.\\
\underline{Correspondence of solutions} \\
{\bf Uniform interaction  $J_{jk}=J_0/N$}\\ 
In this case, $l=1, a_1=J_0, q_{1,j}=1$.
The order parameter is defined as
\bea
R e^{i \Theta} &= & R_\rmR + i R_\rmI = \frac{1}{N} \sum_j e^{i\phi_j}.
\eea
The Hamiltonian is $H=-J_0N(R_\rmR^2+R_\rmI^2)/2 + C$.
 For the  phase oscillator network, this is the Kuramoto model.
$ A_j= \Xi_j =J_0 R $ and $\alpha_j= \phi_j ^0 = \Theta$ follow from 
Eq. (\ref{alpha}). From Eq. (\ref{sce.barg}), 
 the SCE for the order parameter $R$ is 
\bea
R &=&  J_0 R \bar{g}_{\omega_0,\sigma}(J_0 R).
\label{sce.u}
\eea
On the other hand, for the XY model, from Eq. (\ref{spe.baru})  we obtain the SCE as 
\bea
R  &= & J_0 R \bar{u} _\beta (J_0 R).
\label{spe.u}
\eea
 Let us define $v(x)$ and $\bar{J}_0$ as follows:
\bea
v(x) &=& \left\{
\begin{array}{ll}
\bar{q}_{\omega_0,\sigma}(x)/\bar{q}_{\omega_0,\sigma}(0)
=4/(\pi g(\omega_0)) \\
 \times  \int_0 ^{\pi/2} d \psi g(\omega_0+ x \sin \psi)\cos^2 \psi, 
& \mbox{ \rm Oscillator,} \\
\bar{u}_{\beta}(x)/\bar{u}_{\beta}(0)
= 2 u(\beta x),  
& \mbox{ \rm XY model,}
\end{array}
\right.
\eea
\bea
\bar{J_0} &=& \left\{
\begin{array}{ll}
 \bar{q}_{\omega_0,\sigma}(0)J_0
=\pi g(\omega_0) J_0/2, 
& \mbox{ \rm Oscillator,} \\
\bar{u}_{\beta}(0)J_0 =\beta J_0 /2,
& \mbox{ \rm XY model.}
\end{array}
\right.
\eea
We put $x=J_0R$ and $\xi=1/\bar{J_0}$.
Then, SCE and SPE become
\bea
\xi &=& v(x). 
\label{eq.for.1}
\eea
Since $v(0)=1$ and $v(x)$ decreases monotonically to 0 
as $x$ increases from 0 to infinity,
  for any $\xi \in (0, 1]$, eq. (\ref{eq.for.1}) has 
the unique solution. Thus, there is one to one correspondence
between  solutions of the 
SPE and SCE.  The critical point is $\bar{J}_0=1$.
\\
{\bf Mexican-hat type interaction}\\
Now, let us consider the system on a circle.
 We study the Mexican-hat type interaction 
which is given by
\bea
J_{jk} &=& J_0/N + (J_1/N) \cos (\theta_j - \theta_k),
\eea
where $\theta_j$ is the coordinate on the circle, 
$ \theta_j = 2 \pi j /N, \ j=0, 1,  \cdots, N-1.$
The order parameters other than $R$ are defined as
\bea
 R_{1c}e^{\rmi \Theta_{1c}} &= & \frac{1}{N} \sum_j \cos \theta_j e^{\rmi \phi_j}, \\
R_{1s}e^{\rmi \Theta_{1s}} &= & \frac{1}{N}\sum_j \sin \theta_j e^{\rmi \phi_j}. 
\eea
We define $R_1=\sqrt{R_{1c} ^2+R_{1s} ^2}$. 
In order to indicate the location we use $\theta$ instead of $j$. 
There are three non-trivial solutions, the uniform (U) solution ($R>0, R_1=0$), 
the spinning (S) solution ($R=0, R_1>0$) and the pendulum (Pn) solution ($R>0, R_1>0$).
 See Ref. 8 for details.
The uniform solution is equivalent to the solution of the Kuramoto model.
Now, let us study the stable spinning solution.
 $R_{c}=R_{s}$ follows.  We define $\bar{J}_1$ as
\bea
\bar{J}_1 &=& \left\{
\begin{array}{ll}
 \bar{q}_{\omega_0,\sigma}(0)J_1
=\pi g(\omega_0) J_1/2, 
& \mbox{ \rm Oscillator}, \\
\bar{u}_{\beta}(0)J_1 =\beta J_1 /2,
& \mbox{ \rm XY model}.
\end{array}
\right.
\eea
We put $x=J_1 R_{1c}$ and $\eta=1/\bar{J}_1$.
The SCE and SPE become
\bea
\eta &=& v(x)/2.
\label{eq.spin}
\eea
Therefore, for any $\eta \in (0, 1/2]$,
there exists the unique solution of (\ref{eq.spin}).
 Thus, the solutions for the SCE and SPE correspond uniquely. 
The critical point is $\bar{J}_1=2$.  \\
Next, we study the stable pendulum solution.
We define 
$ x=J_0 R$ and $y=J_1 R_{1}$.
The SCEs and SPEs become
\bea
\xi &=& F(x,y) = \bra v(\Lambda(x,y, \theta)) \ket,  \label{pn.eq.1}\\
\eta &=& G(x,y) =\bra v(\Lambda(x,y, \theta)) \cos ^2 \theta \ket, \label{pn.eq.2}\\
\bra B \ket &=& \frac{2}{\pi} \int_0 ^{\pi/2} d \theta B,
\eea
where $\Lambda(x,y, \theta) = \sqrt{x^2+y^2 \cos ^2 \theta}$.
We get
\bea
F(x,0)&=&v(x), \\ 
F_y(x,y)&=& \partial F(x,y) / \partial y
<0, \mbox{  for } x \ge 0, \  y>0.\\
 \lim_{y \to \infty}F(x,y)&=&0  \mbox{ for } x \ge 0.
\eea
Thus, for fixed $x \ge 0$, $F(x,y)$ is a decreasing function of $y$.
For $\xi \in (0, 1]$, 
there is the unique solution of  $v(x)=\xi$.
 We denote it by $x_0=v^{-1}(\xi)$. Note that  $x_0=0=v^{-1}(1)$.
 Therefore, for any $x \in [0, x_0]$ there exists the unique solution 
of (\ref{pn.eq.1}),
\bea
 y&=&y(\xi, x). \label{sol.eq.1}
\eea
We have relations $y(\xi, x_0)=0$ and $y(1, 0)=0$. 
Substituting eq. (\ref{sol.eq.1}) into eq. (\ref{pn.eq.2}) we get
\bea
\eta &=&G(x, y(\xi, x)). 
\label{pn.eq.2-2}
\eea
It is proved that $ G(x, y(\xi, x))$ is a strictly increasing function of $x$
 for $x>0$.  
Since $y(\xi, x)$ exists for $0 \le x \le x_0$,
when $G(0, y(\xi, 0))\le \eta \le G(x_0, y(\xi, x_0)) $,
the solution $x(\xi, \eta)$ of eq. (\ref{pn.eq.2-2}) uniquely exists.
$y(\xi, 0)$ is determined by
\bea
\xi &=& F(0, y(\xi, 0))=\bra v (y(\xi, 0) \cos  \theta ) \ket.
\eea
On the other hand, because of $v(x_0)=\xi$,
$G(x_0, y(\xi, x_0)) $ is given by
\bea
G(x_0, y(\xi, x_0))&=& G(x_0, 0)=\bra v(x_0)\cos ^2 \theta \ket =\xi /2.
\eea
Thus, defining $\eta_0 (\xi) \equiv G(0, y(\xi, 0))$,
for $\eta_0 (\xi) \le \eta \le \xi/2$, the solution of 
eq. (\ref{pn.eq.2-2}) uniquely exists.
 The condition $\eta \le  \xi/2$ implies $J_1 \ge 2 J_0$,
and this is the condition that the Pn solution emerges from the U solution
\cite{uezu.osc,kimoto.xy}.
 On the other hand, the condition $\eta=\eta_0(\xi)$ is considered to be
 that the stable Pn solution becomes unstable 
and then disappears by merging with the unstable S solution.\\
\underline{Application} \\
Now, let us consider an application of the correspondence between the two models.  
To obtain non-trivial results, we study the following associative memory type interaction:
\bea
J_{jk}&=& \frac{J}{N} \sum_{\mu=1}^p \xi_i ^\mu \xi_j ^\mu,
\eea
where  ${\mbox{\boldmath $\xi$}}^{\mu} 
=(\xi^{\mu}_{1},\xi^{\mu}_{2},...,\xi^{\mu}_{N})$
is the $\mu$th  pattern ($\mu=1, 2, \cdots, p$).  
That is, $a_\mu=J, q_{\mu, j}= \xi_j ^\mu .$ 
 We assume that $p \ll N$ and 
$\xi^{\mu}_i$ take values of  $\pm 1$, and correlate with each other as follows:
\bea
\bra \xi_i ^\mu  \xi_j ^\nu  \ket &=& \biggl( a+(1-a)  \delta_{\mu,\nu} \biggr) \delta_{i,j}.
\eea
The XY model with this interaction has a peculiar solution, that is, 
there exists one-parameter family of solutions of the SPEs\cite{kimoto.etal}.
We call this solution the continuous solution.
Here, we derive the SPEs of this solution.
We introduce  sublattices $\Lambda_l$ in which the following holds:
\bea
(\xi_i ^1, \xi_i ^2, \cdots,  \xi_i ^p)&=&(\eta_l ^1, \eta_l ^2, \cdots,  \eta_l ^p)
\ \ \mbox{ for } i \in \Lambda_l,\\
\eta^\mu _{l+2^{p-1}} &= &-\eta^\mu _{l}, \ l=1,2,\cdots, 2^{p-1}. 
\eea
The number of elements in $\Lambda_l$, $|\Lambda_l|$, is 
$\displaystyle |\Lambda_l|=N/2^p \ (l=\!1, 2, \cdots,  2^p)$.
Order parameters are defined as 
\bea
R_\mu e ^{i \Theta_\mu} &=& R_{\mu \rmR} + R_{\mu \rmI} 
= \frac{1}{N}\sum_{j=1}^N \xi_j ^{\mu} e^{\phi_j}, \ \mu=1,\cdots,p.
\eea
The Hamiltonian is rewritten as 
\bea
H&=&  - \frac{N}{2} J \sum_{\mu=1}^p R_{\mu} ^2 + C.
\eea
From Eq. (\ref{spe}) the SPEs become
\bea
 R_{\mu } e^{i\Theta_\mu}&=& \beta J 
\bra u(x_j)\sum_{\nu}  (R_{\nu \rmR}+ i R_{\nu \rmI}) \xi_j ^{\nu} \xi_j^\mu \ket,\\
\Xi_j &=& \sqrt{(\sum_j R_{\mu \rmR} \xi_j ^\mu)^2 +(\sum_j R_{\mu \rmI} \xi_j ^\mu)^2 },
\eea
where $\Xi_j$ is  redefined as $\Xi_j$ in Eq. (\ref{def.xi}) divided by $J$.
$ x_j = \beta J \Xi_j$, and $\bra \cdot \ket $ implies the average over
$\{\xi_j\}$.
We define the probability $P_l$ 
that $\{ \xi_i ^\mu \}$ take values in the $l$th sublattice.
The SPEs are rewritten as
\bea
 R_{\mu \rmR}&=& \beta J \sum_{\nu} c_{\mu \nu} R_{\nu \rmR},
\label{R-1}\\
 R_{\mu \rmI}&=& \beta J \sum_{\nu} c_{\mu \nu} R_{\nu \rmI},
\label{R-2}\\
 c_{\mu \nu} & = & 2 \sum_{l=1}^{2^{p-1}} P_l u_l \eta_l ^\mu \eta_l ^\nu =c_{\nu \mu},
\label{c}
\eea
where $u_l = u(x_l), \ x_l = \beta J \Xi_l$, and 
 $\Xi_l$ is $\Xi_j$ evaluated at the $l$th sublattice.
By defining $R=\sqrt{\sum_\mu R_\mu ^2}$, we obtain additional equations from 
 Eqs. (\ref{R-1}) and (\ref{R-2}) as
\bea
R^2 &=& \frac{1}{2^{p-1}}  \sum_{l=1} ^{2^{p-1}}\biggl( \frac{x_l}{\beta J}
\biggr) ^2,
\label{R}\\
R^2 &=& \frac{2}{\beta J} \sum_{l=1}^{2^{p-1}} P_l u_l x_l^2.
\label{R2}
\eea
The SPEs of the continuous solution are
\bea
c_{\mu \nu} = \delta_{\mu \nu} \frac{1}{\beta J}. 
\label{cont.sol}
\eea
Hereafter, we study the case $a=0$ for simplicity.
For $p=2$, from (\ref{cont.sol}), $u_1=u_2=1/(\beta J)$,  and thus   $x_1 = x_2$ follow.
From (\ref{R}), we obtain $R=\Xi_1= x_1/(\beta J)$.
 Thus, the SPE is rewritten as
\bea
\bar{u}_\beta(J R) & =  & 1/J,
\label{spe.aso}
\eea
which determines $R$.
 Thus, the continuous solution is given by
\bea
&&  0 \le R_1 \le R, \ R_1 ^2+R_2 ^2 = R^2.
\label{condR}
\eea
This implies that any point $(R_1, R_2)$ on the circle 
 connecting two points representing patterns $\vecxi ^1$ and $\vecxi ^2$ is a solution.
For general $p$, the circle connecting any two points representing  patterns  
 $\vecxi ^\mu$ and $\vecxi ^\nu$ is a continuous solution.
We performed Markov chain Monte Carlo (MCMC) simulations for $p=2$ and 3.
We show the result for $p=2$ in Fig. 1. 
We note that the trajectories of $R_1$ and $R_2$ wander but 
 $R$ is almost constant. As seen from Fig 1. (c), 
theoretical and numerical results agree quite well.
\begin{figure}[ht]
\begin{picture}(120,115)
\put(-70,115){(a)}
\put(-140,5){\includegraphics[width=5cm,clip]{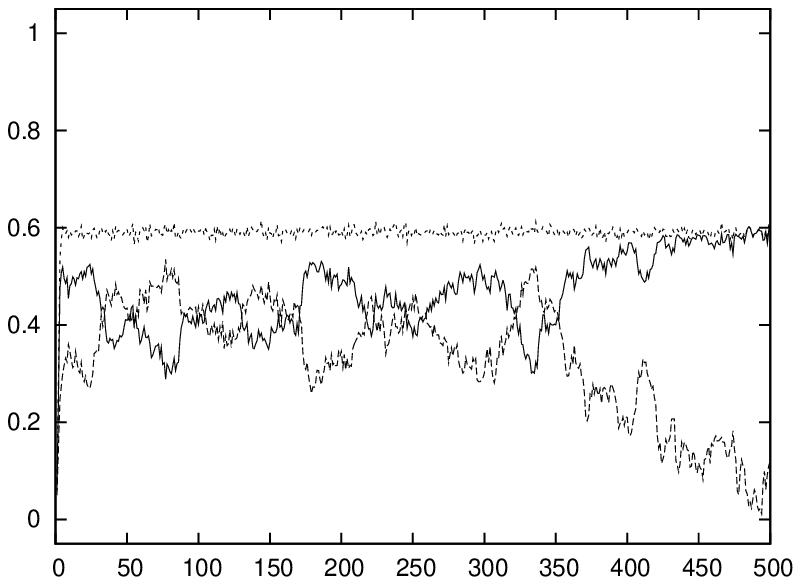}}
\put(-65,70){$R$}
\put(-20,50){$R_1$}
\put(-35,20){$R_2$}
\put(-110,-5){time [$\times 100$ mcs]}
\put(80,115){(b)}
\put(20,5){\includegraphics[width=4cm,clip]{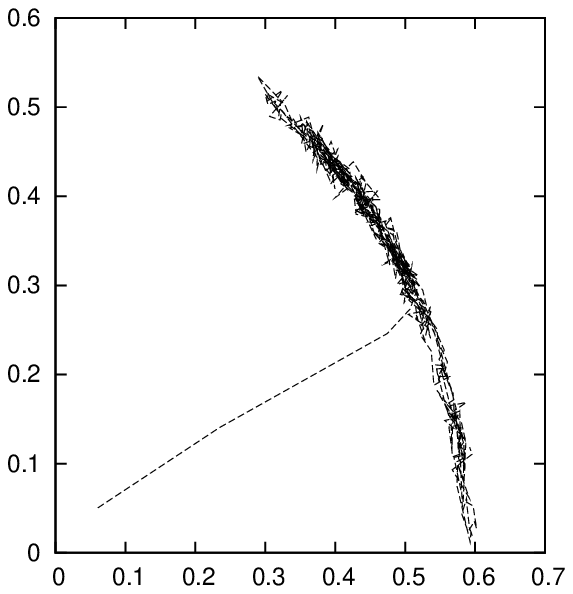}}
\put(15,55){$R_2$}
\put(80,-5){$R_1$}
\put(220,115){(c)}
\put(150,5){\includegraphics[width=5cm,clip]{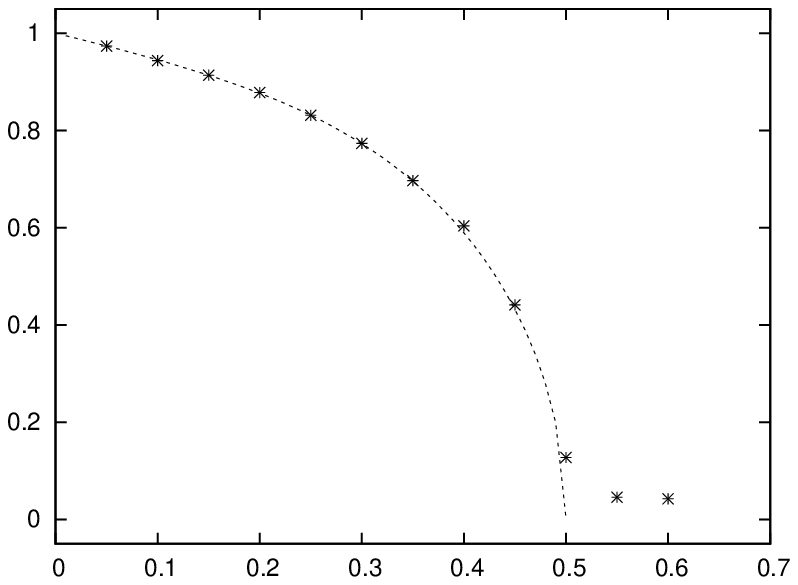}}
\put(145,55){$R$}
\put(220,-5){$T$}
\end{picture}
\caption{XY model. $J=1$.
(a), (b)  Time series of  $R_1, R_2$, and $R$,  
 and trajectories in $(R_1, R_2)$ space obtained by Monte Carlo simulations.  $N=10^4$. $T=0.4$.
(c) $T$ dependence of  $R$.   Curve: theoretical results,  symbols: numerical results.  $N=10^4$.
}
\label{fig.1}
\end{figure}
%
%
%
%
Next, let us study the phase oscillator network with the same interaction.
  The SCE is immediately obtained by the correspondence of the integral kernels,
\bea
\bar{q}_{\omega_0,\sigma}(J R )& = & 1/J.
\eea  
This is simply the SCE of the Kuramoto model.  Since we  have the same relation as
Eq. (\ref{condR}), we also obtain the continuous solution.
 We performed numerical integrations of Eq. (\ref{eq:evolution}) for $p=2$ and 3.
 We took a Gaussian distribution with a mean 0 and a standard deviation $\sigma$ for 
 $g(\omega)$.
We used the Euler method with the time increment $\Delta t = 0.1$.
See Fig. 2. 
There should exist continuous stationary states, but instead, we found  
 a noisy  limit cycle oscillation.   The reason for this is considered as follows:
In the derivation of the SPE (\ref{sce}), the desynchronized oscillators
 do not contribute. However, in numerical simulations,
 the desynchronized oscillators contribute to the dynamics  because $N$ is
 finite. Since the continuous stationary states easily move to
the marginally stable  direction by perturbations, the trajectories
 move on the manifold of $R_1^2 + R_2 ^2 =R^2$.  
 This is confirmed by Fig. 2(b).  Figure 2(c) on the  $\sigma$ dependence of $R$
 shows fairly good agreement
between the theoretical results for the continuous solution 
 and the numerical results for the limit cycle. 
\begin{figure}[ht]
\begin{picture}(120,115)
\put(-70,115){(a)}
\put(-140,5){\includegraphics[width=5cm,clip]{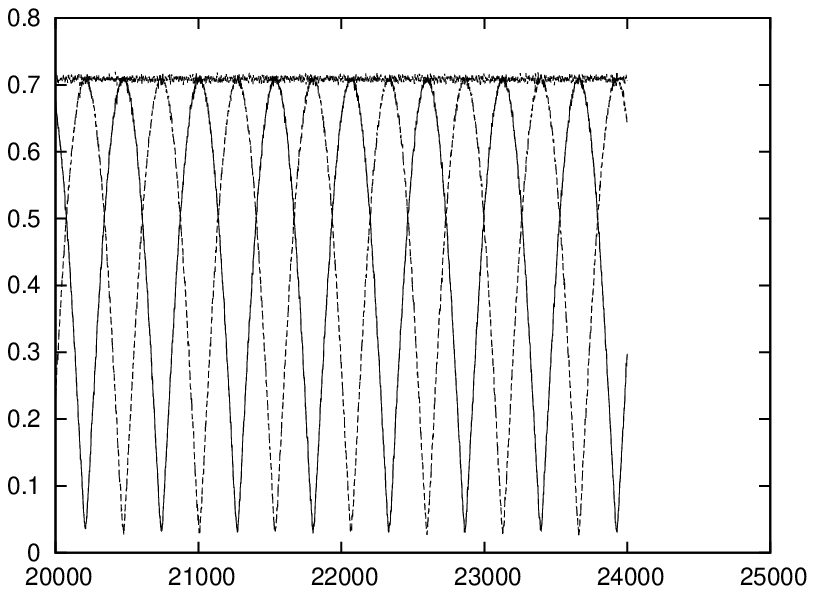}}
\put(-25,90){$R$}
\put(-25,30){$R_1$}
\put(-25,70){$R_2$}
\put(-65,-5){$t$}
\put(80,115){(b)}
\put(20,5){\includegraphics[width=4cm,clip]{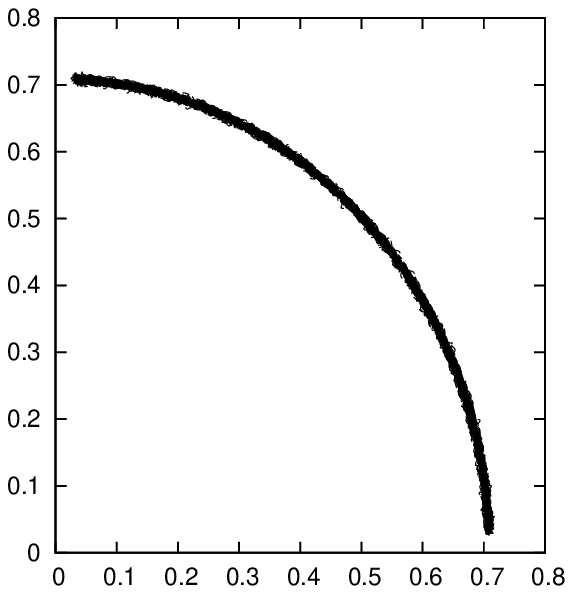}}
\put(15,55){$R_2$}
\put(80,-5){$R_1$}
\put(220,115){(c)}
\put(150,5){\includegraphics[width=5cm,clip]{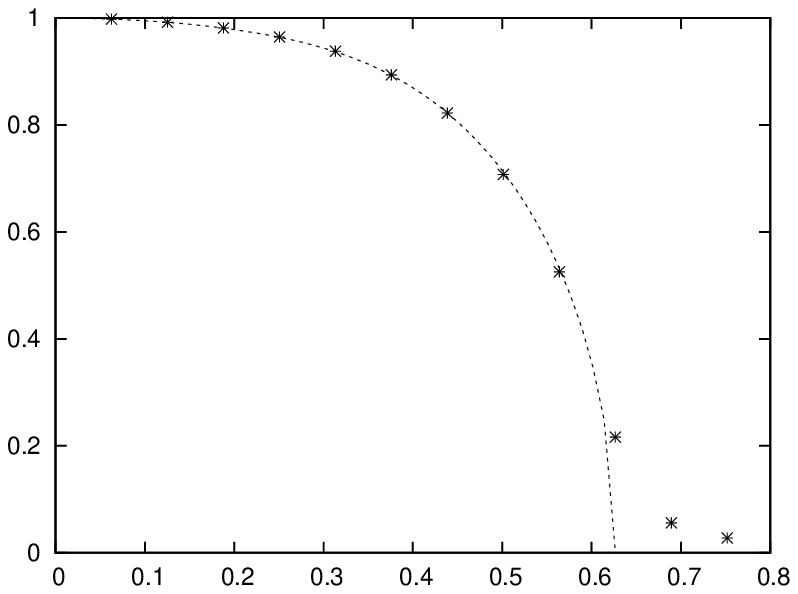}}
\put(145,55){$R$}
\put(220,-5){$\sigma$}
\end{picture}
\caption{Phase oscillator network. $J=1$
(a), (b)   Numerical results of time 
series of  $R_1, R_2$, and $R$,   and  
 trajectories in $(R_1, R_2)$ space obtained by the Euler method.  
$N=10^4, \sigma=\sqrt{\frac{\pi}{2}} T$ with $T=0.4$.
(c) $\sigma$ dependence of  $R$.   
Curve: theoretical results, symbols: numerical results.
$N=10^4$.
}
\label{fig.2}
\end{figure}

In summary, we studied the correspondence between 
the phase oscillator networks and the classical XY models
 with  the same infinite-range interactions.
Assuming a class of interactions, 
we found the correspondence between the integration kernel 
of the SCEs for the oscillator network and that of the SPEs for the XY model. 
 We found a sufficient condition that the integration kernel of
 the SCEs for the oscillator network has the same feature as that of
the SPEs for the XY model. 
That is, the probability density of the natural frequency distribution
  is  one-humped. 
 Furthermore, we found that  the quenched randomness 
(distributed natural frequencies) corresponds to thermal noise.
 To study the correspondence of solutions in both models, 
we  investigated the uniform interaction and the Mexican-hat type interaction   on a circle.
 We proved that
the solutions  uniquely correspond  in both models. 
As an application of the correspondence, we 
studied  the associative memory type interaction,
for which the XY model has a peculiar one-parameter family
 of solutions called the continuous solution.
We found that the continuous solution is not stable for the oscillator network,
and instead a noisy limit cycle appears,  which lies on the manifold that
the continuous solutions exist.
We consider that this is caused by the desynchronized
oscillators and  is a  finite size effect.

 When $g(\omega)$ is the uniform distribution which  is not one-humped,
 we can prove the one to one correspondence of solutions for some interactions in both models.
 This will be reported elsewhere.

For the interactions studied in this paper, there exist several types of
solutions, and we found that  the stabilities of the corresponding
 solutions in both models  are the same except for the continuous solution.
In order to study the stability of a solution for the oscillator network, we have to
 derive the evolution equations for order parameters and it is 
 a very  difficult problem to   solve\cite{ott.antonsen}.
The correspondence of stabilities of solutions in  both models is left as a future problem.

\end{document}